\documentclass[structabstract]{aa} 
\usepackage{graphicx}
\usepackage{txfonts}
\begin{document}
%

   \title{Radio continuum emission from knots in the DG Tau jet}

   \author{L. F. Rodr\'\i guez
          \inst{1,2},
	  R. F. Gonz\'alez
	  \inst{1},
	  A. C. Raga
	  \inst{3},
	  J. Cant\'o
	  \inst{4},
	  A. Riera
	  \inst{5},
	  L. Loinard 
	  \inst{1,6},
	  S. A. Dzib
	  \inst{1}
	  \and
	  L. A. Zapata
	  \inst{1}
                 }

\institute{Centro de Radioastronom\'\i a y Astrof\'\i sica, Universidad Nacional 
Aut\'onoma de M\'exico, A. P. 3-72, (Xangari), 58089 Morelia, Michoac\'an, M\'exico\\ 
\email{l.rodriguez@crya.unam.mx} \\
\and
Astronomy Department, Faculty of Science, King Abdulaziz University, 
P.O. Box 80203, Jeddah 21589, Saudi Arabia\\
\and
Instituto de Ciencias Nucleares, Universidad Nacional Aut\'onoma de M\'exico, Apdo. 
Postal 70-543, CP. 04510, D. F., M\'exico\\
\and
Instituto de Astronom\'\i a, Universidad Nacional Aut\'onoma de M\'exico, Apdo.
Postal 70-264, CP. 04510, D. F., M\'exico\\
\and
Departament de F\'\i sica i Enginyeria Nuclear, EUETIB, Universitat Polit\`ecnica de Catalunya, 
Compte d'Urgell 187, 08036 Barcelona, Spain\\
\and
Max-Planck-Institut f\"ur Radioastronomie, Auf dem H\"ugel 69, 53121 Bonn, Germany\\
   }

   \date{Received , 2011; accepted , }


  \abstract
    {HH~158, the jet from the young star DG Tau, is one of the few sources of
    its type where jet 
    knots have been detected at optical and X-ray wavelengths.}
     {To search, using Very Large Array observations of this source, radio
     knots and if detected, compare them with
     the optical and X-ray knots. To model the 
     emission from the radio knots.}
  {We analyzed archive data and also obtained new Very Large Array observations of this source, 
  as well as an 
  optical image, to measure the present position of the knots. We also modeled the 
  radio emission from the knots
  in terms of shocks in a jet with intrinsically time-dependent ejection velocities.}
     {We detected radio knots in the 1996.98 and 2009.62 VLA data. These radio knots are,
     within error, coincident with optical knots. We also modeled satisfactorily
     the observed radio flux densities as
     shock features from a jet with intrinsic variability. 
     All the observed radio, optical, and X-ray knot
     positions can be intepreted as four successive knots, ejected with a period
     of 4.80 years and traveling away from the source with a velocity of 198 km s$^{-1}$
     in the plane of the sky.}
     {The radio and optical knots are spatially correlated and 
     our model can explain the observed radio flux densities. However, the X-ray knots do
     not appear to have optical or radio counterparts and their nature remains poorly
     understood.}

   \keywords{ISM: jets and outflows -- Herbig-Haro objects -- radio-continuum: ISM --
                 stars: individual: DG Tau
               }

\authorrunning{Rodr\'\i guez et al.}
\titlerunning{Radio continuum observations of the DG Tau jet}
   \maketitle
%

\section{Introduction}
HH 158, the jet from DG Tauri was first reported by Mundt \& Fried (1983), who
presented an H$\alpha$ image showing a well defined HH knot
at $\sim 8''$ to the SW of the star
connected to DG Tau itself by a faint bridge. High resolution
spectroscopy of this jet was presented by Mundt et al. (1983) and
Solf \& B\"ohm (1993). In the latter paper, the jet emission was traced
to within $\sim 0\rlap.{''}2$ from DG Tau.

DG Tau is located in the sky approximately in between 
the L1495 region and the star HP Tau. There are accurate distance
determinations from very long baseline interferometry geometric parallax to both 
the L1495 region (131.5 pc; Torres et al. 2007; 2009)
and HP Tau (161 pc; Torres 2009). Here we adopt for DG Tau a distance
of 150 pc, intermediate to those of L1495 and HP Tau.

The proper motions
of the knots observed up to distances of $\sim 10''$ were derived
by Eisl\"offel \& Mundt (1998), using several frames obtained
over a $\sim 7$ yr time span, giving velocities of
$\sim 150$ km s$^{-1}$  (assuming a distance of 150 pc).
Comparing adaptive optics
images obtained with a $\sim 2$ yr time base, Dougados et al.
(2000) obtained proper motion velocities of $\sim 200$ km s$^{-1}$
for the knots within $\sim 6''$ from the source. These velocities
imply a dynamical timescale of $\sim 40$ yr for HH 158 at that epoch.

McGroarty \& Ray (2004) suggested
that two groups of HH knots (HH 702 to the SW and HH 830 to the NE)
at angular distances of $\sim 11'$ from DG Tau might be associated
with the same outflow. However, McGroarty et al. (2007) showed
that proper motion measurements only support the possibility
of HH 702 being associated with the outflow from DG Tau 
(the HH 830 knots have proper motions which are not aligned with
the outflow axis). The HH 702 knots have proper motions of
$\sim 240$ km s$^{-1}$,
resulting in a dynamical timescale of $\sim 2000$ yr.

Spectroscopic data with 2D angular resolution (but with lower spectral
resolution) of the region around DG Tau were presented by Lavalley
et al. (1997, 2000). These observations show that some of the ejections
from DG Tau have a remarkable, bow shock-like morphology, and also
show in a clear way the presence of a faint counterjet
to the NE (marginally
seen in some of the previous observations).

Interestingly, the only optical spectrophotometric study of HH 158
with an extended wavelength coverage
appears to be the one of Cohen \& Fuller (1985). The spectrum
described by these authors covers from $\sim$4000 to $\sim$7000\AA,
and shows (reddening corrected) ratios of [O III] 5007/H$\beta$=0.28,
[O I] 6300/H$\alpha$=0.30, [N II] 6548+83/H$\alpha$=0.64
and [S II] 6716+31/H$\alpha$=0.69. These observed [O III]/H$\beta$ and
[S II]/H$\alpha$ ratios identify HH 158 as a high excitation
HH object (see Raga et al. 1996).

More recent work has focussed on the low excitation lines.
The paper of Solf \& B\"ohm (1993, mentioned above) and
the more recent papers of Bacciotti et al. (2000)
and Coffey et al. (2007, 2008, who
present red and near UV STIS spectra), and Pyo et al. (2003, who
study the [Fe II] 1.644 $\mu$m emission) do not cover the blue
region of the spectrum. Because of this, the [O III] 5007 emission
reported by Cohen \& Fuller (1985) has not been re-observed.

Herczeg et al. (2006) obtained FUV spectra of DG Tauri, reporting
the detection of fluorescent H$_2$ lines, and the non-detection
of lines like CIV 1549, which would be expected in the spectrum
of a high excitation HH object. This non-detection possibly indicates
that the high excitation emission region detected by Cohen \& Fuller
(1985) might be absent two decades later, or that it lies outside
the region sampled by the slit in the STIS spectrum of Herczeg et
al. (2006).

The high excitation nature of HH 158 has been confirmed by the
somewhat surprising detection of extended X-ray emission along
the outflow (G\"udel et al. 2005, 2007, 2008, 2011; Schneider \& Schmitt
2008; G\"unther et al. 2009). These observations show an X-ray knot at
$\sim 5''$ from DG Tau, along the direction of the HH 158 flow.

We have obtained multiepoch VLA images and a red [S II] image of the region
around DG Tauri, in order to explore the current morphology
of the outflow, and to be able to relate the X-ray
emission to the structures observed at other wavelengths.

The base of the HH 158 flow was detected in the VLA
observations of Cohen \& Bieging (1982) and Bieging et al. (1984).
In the present paper we present new VLA observations made at 3.6 cm
in 1994, 1996, and 2009. We use these data, together
with the knot positions measured over 
the past 20 years at optical,
IR and X-ray wavelengths in order to derive a kinematical model
for the evolution of the HH 158 outflow.

The paper is organized as follows. In section 2, we describe
the new and archival optical and radio continuum observations.
In section 3, we present the sequence of radio continuum maps, and
the red [S II] image. In section 4, we derive a simple kinematic
model for the time-evolution of the knot structure of HH 158
over the past 20 years, and a model for the free-free
continuum produced in variable ejection velocity jets. Finally, the results are summarized
in section 5.

\section{Radio and optical observations}

\subsection{Radio continuum observations}

\begin{figure}
   \centering
   \includegraphics[width=8cm]{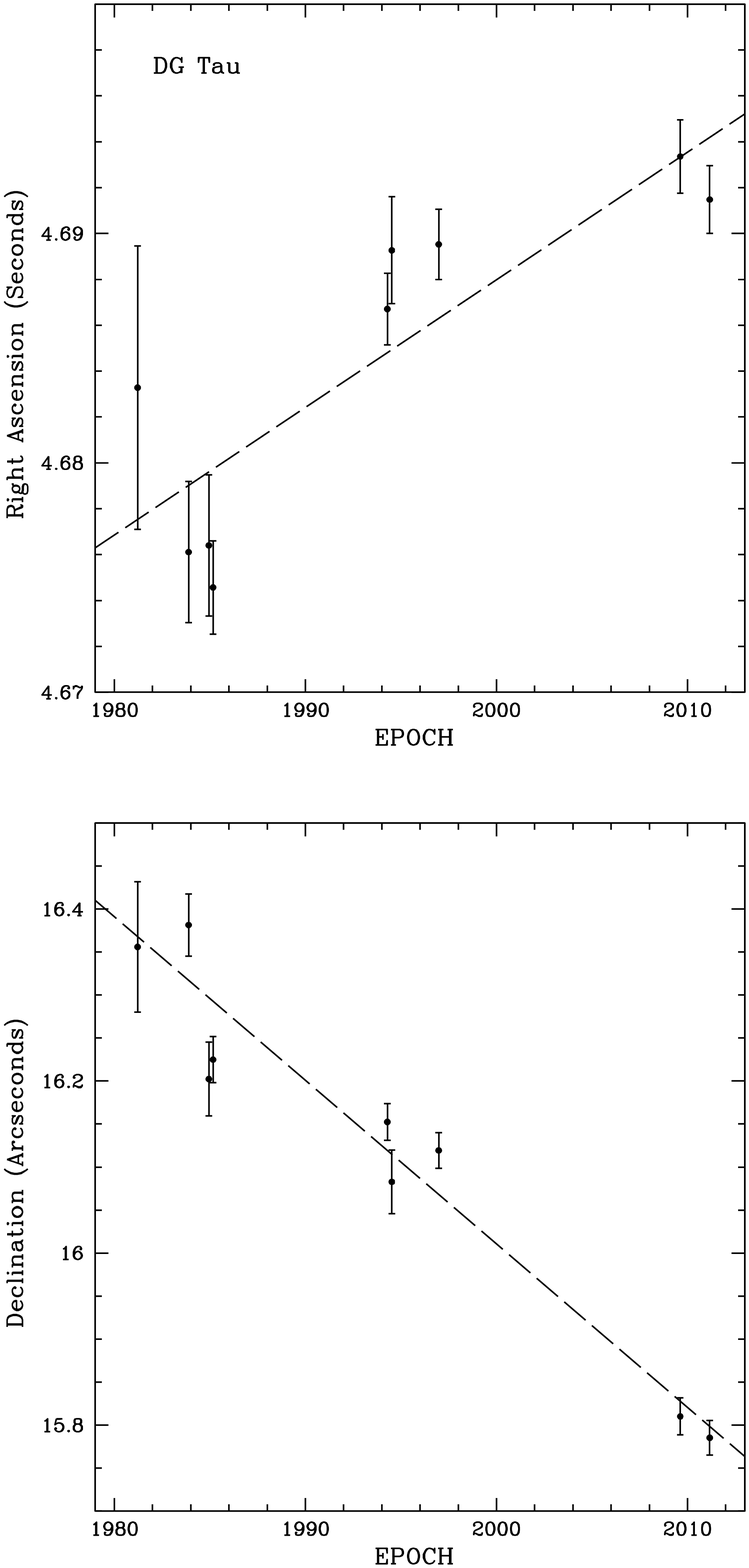}
                 \caption{\small Position of the radio emission of DG Tau
                 as a function of time from VLA and EVLA data. The right ascension (top) is with
                 respect to $04^h~22^m$
                 and the declination (bottom) is with respect to $+26^\circ~06'$.
                 The dashed lines are least-squares fits to the data that
                 give the proper motions discussed in the text.}
\label{propermotions}
\end{figure}

Since the objects in Taurus are known to exhibit relatively large proper motions
(tens of mas yr$^{-1}$, i. e. Loinard et al. 2007; Torres et al 2007) and our observations cover large
time intervals (15 years), we considered it necessary to first determine the
proper motions of DG Tau to correct the positions
and understand better any possible changes.
For this proper motion determination we used the three epochs observed by us
and described in detail below, as well as five additional observations taken from the VLA
archive. Finally, we also included an EVLA (Expanded Very Large Array)
observation taken in 2011 February 26 at an average wavelength
of 5.3 cm (two 1 GHz bandwidths centered at 4.56 and
7.43 GHz) in the B configuration, as part of 
the Gould's Belt Distance Survey (Loinard et al. 2011).
The full set of 9 observations covers 
about 30 years. The radio positions of DG Tau as a function of time are shown in
Figure \ref{propermotions}. The proper motions of the star derived from the
fits shown in this Figure are:

\begin{eqnarray}
\mu_\alpha\cos \delta  & = &  +7.5 \pm 0.9 \mbox{~mas~yr$^{-1}$}\nonumber\\
\mu_\delta  & = &  -19.0 \pm 0.9 \mbox{~mas~yr$^{-1}$.} \nonumber
\end{eqnarray}

These proper motions are consistent within 2-$\sigma$ with those reported
by Ducourant et al. (2005) from optical observations.
Since our proper motion determination
is about a factor of two more accurate than that of Ducourant et al. (2005),
we adopt it and use for the position of DG Tau the following values:

\begin{eqnarray}
\alpha(2000) = 04^h~27^m~04\rlap.{^s}6880 + 0\rlap.{^s}00055  \times (epoch - 2000.0) \nonumber\\
\delta(2000) = +26^\circ~06'~16\rlap.{''}011 - 0\rlap.{''}0190 \times (epoch - 2000.0), \nonumber
\end{eqnarray}

\noindent where $epoch$ is the epoch given in decimal years.
In the images discussed below we use this position corrected for the epoch
of the observation.

The Very Large Array (VLA) observations made by us in the continuum at 3.6 cm
were taken in three epochs spanning 15 years. The parameters of these observations are summarized
in Table 1. The data were edited
and calibrated using the software package Astronomical Image Processing System (AIPS) of 
the USA National Radio Astronomy Observatory (NRAO).

\begin{table*}[htbp]
\small
      \caption{VLA Observations}
        \begin{center}
            \begin{tabular}{lccccccc}\hline\hline
  &         & Frequency & On-source & Number of & Phase & Bootstrapped & Beam  \\
  Epoch/Configuration  & Project & (GHz)& Time (min) & Antennas &
  Calibrator & Flux Density (Jy) & Angular Size \\
  \hline
  1994 Apr 16 (1994.29)/A & AR277   & 8.46 & 206 & 18 & 0403+260 & 0.739$\pm$0.002 &
  $0\rlap.{''}19 \times 0\rlap.{''}16;~\ $-$68^\circ$  \\
  1996 Dec 24 (1996.98)/A & AR277   & 8.46 & 631 & 14 & 0403+260 & 0.691$\pm$0.002 &
  $0\rlap.{''}25 \times 0\rlap.{''}24;~\ +19^\circ$  \\
  2009 Aug 13 (2009.62)/C & AR694   & 8.46 & 414 & 26 & 0403+260 & 2.34$\pm$0.02 &
  $3\rlap.{''}07 \times 2\rlap.{''}93;~\ $-$15^\circ$  \\
  \hline\hline
      \label{tab:1}
  \end{tabular}
    \end{center}
\end{table*}

\begin{figure}
   \centering
   \includegraphics[width=8cm]{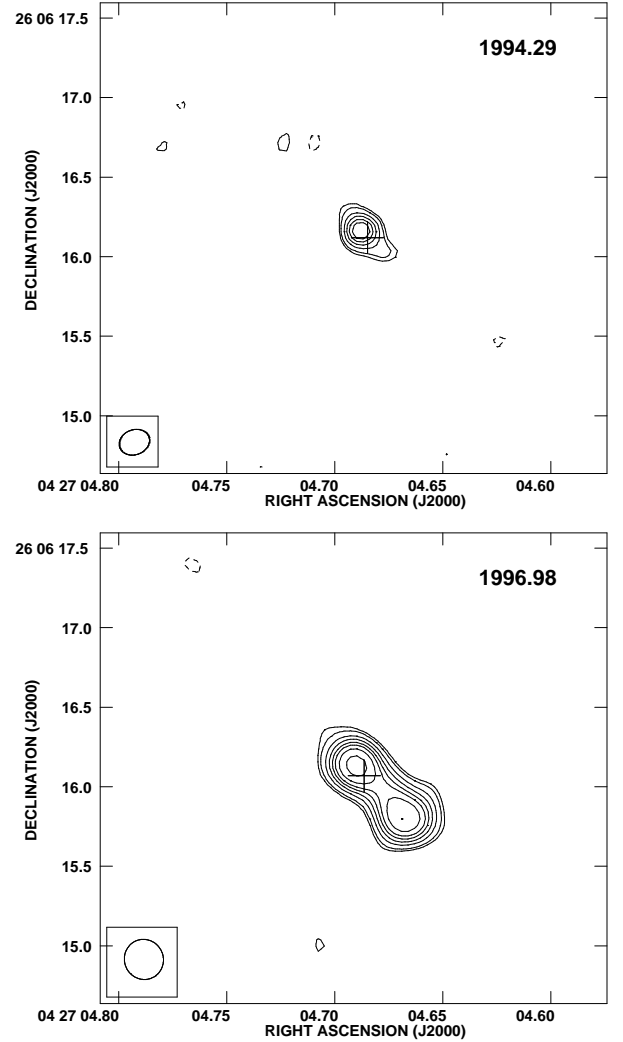}
   \caption{\small VLA contour images of the 3.6 cm emission associated
   with DG Tau for the epochs 1994.29 (top) and 1996.98 (bottom).
   The images were made with the weighting parameter ROBUST = 0.
   The contours are -3, 3, 4, 6, 8, 10, 12, 15, 20, and 25 times
   19 and 15 $\mu$Jy beam$^{-1}$, the rms noises of the 1994.29 and
   1996.98 images, respectively. The crosses mark the position of
   the star, determined as described in the text. The beams are shown in the bottom left corner
   and their dimensions are given in Table 1.}
\label{dgtaux9496}
\end{figure}

In Figure \ref{dgtaux9496} we show the images of the 1994 and 1996 epochs, both
made in the highest angular resolution A configuration. The images are strikingly
different. The 1994 image shows a single component slightly elongated
in the NE-SW direction. The total flux density of this
source is 0.41 $\pm$ 0.04 mJy. In contrast, the 1996 image shows two components,
one is similar to that seen in 1994 but there is a new component
$0\rlap.{''}42 \pm 0\rlap.{''}01$ 
to the SW of the first.
If we assume that this new component is a knot that
was ejected between the two epochs of observation and a distance
of 150 pc, we derive a lower limit of
116 $\pm$ 3 km s$^{-1}$ for its velocity in the plane of the sky.
The flux densities of these
components are 0.44 $\pm$ 0.03 mJy (NE component, corresponding
to the star) and 0.40 $\pm$ 0.03 mJy (SW component, corresponding
to the knot).

\begin{figure}
   \centering
   \includegraphics[width=8cm]{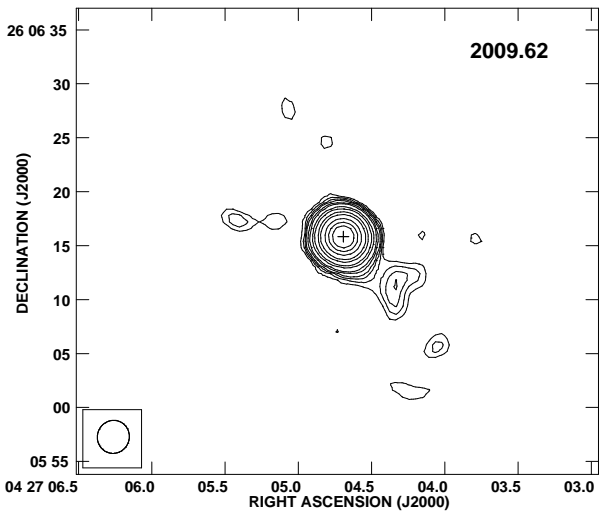}
 \caption{\small VLA contour images of the 3.6 cm emission associated
  with DG Tau for the epoch 2009.62.
  The image was made with the weighting parameter ROBUST = 5 to
  emphasize extended structure.
 The contours are -3, 3, 4, 6, 8, 10, 12, 15, 20, 30, 40, 60, 80, and 100
 25 times 8.8 $\mu$Jy beam$^{-1}$, the rms noise of the
image. The cross marks the position of
   the star, determined as described in the text.
   The beam is shown in the bottom left corner
  and its dimensions are given in Table 1.}
 \label{dgtaux09}
\end{figure}

In addition to the high angular resolution observations made in 1994 and 1996, we recently
made a deep integration of the region also at 3.6 cm but in the configuration C,
that provides an angular resolution of $\sim 3{''}$ to search for extended
components around DG Tau. The image obtained from these observations is
shown in Figure \ref{dgtaux09}. Two components dominate this image: a bright one
with a flux density of 1.47 $\pm$ 0.02 mJy coincident with the star and a
fainter one at $6\rlap.{''}98 \pm 0\rlap.{''}40$ to the SW of the star and with a flux density of
0.15 $\pm$ 0.02 mJy. We attribute this second source to a knot ejected in the past
by DG Tau.

\subsection{Optical observations}

In order to see the present optical structure of the HH~158 outflow,
we have obtained an image of this object in the night of
February 23, 2010. The narrow band image of DG Tau was obtained
at the 2.6m Nordic  Optical Telescope (NOT) of the Roque de Los Muchachos 
Observatory (La Palma, Spain)
using the Service Time mode facility. The image was obtained with 
the Andalucia Faint Object Spectrograph and Camera (ALFOSC) 
in imaging mode. The detector was an E2V 2Kx2K
CCD with a pixel size of 13.5$\mu$m, providing a plate scale 
of 0.19 arcsec pixel$^{-1}$. A [S II] filter (central wavelength 
$\lambda$ = 6725 \AA, FWHM = 60 \AA) was used to obtain an image 
of DG Tau in the [S II] 6716, 6731 \AA~ emission lines. 

Two exposures of 900 seconds each were combined to obtain the final 
image. The angular resolution during the
observations, as derived from the FWHM of stars in the field of 
view, was of 0.9 - 1.0 arcsec. 
The images were processed with the standard tasks of the IRAF reduction 
package. 

\begin{figure}
   \centering
   \includegraphics[width=8cm]{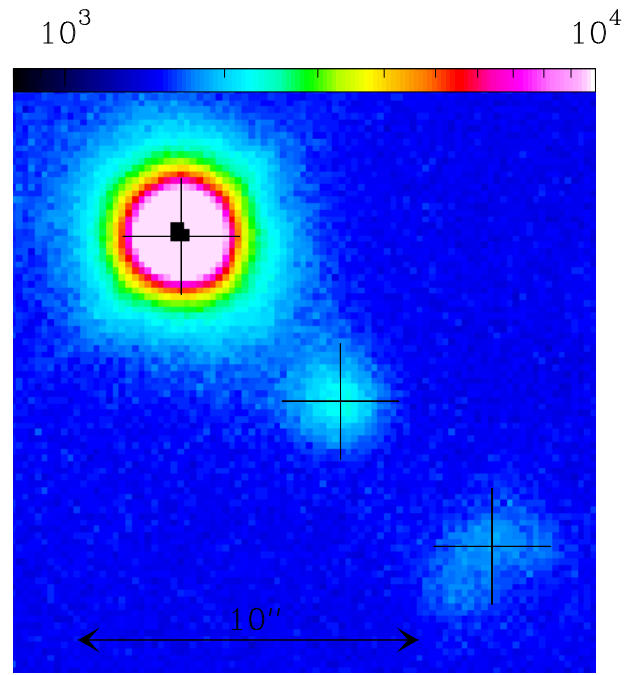}
 \caption{\small Red [S~II] image of the HH~158 outflow.
The three crosses represent the positions deduced from
paraboloidal fits to DG Tau (upper left cross) and
the two knots detected along HH~158. The image is
displayed with a logarithmic scale (given in arbitrary
units by the top bar).}
 \label{dg}
\end{figure}

The [S~II] image is shown in Fig. \ref{dg}. In this figure
we also show the positions found through paraboloidal
fits to the positions of the source and of the two observed knots.
As the star is saturated, we have actually elliminated the central
region, and carried out the fit to an unsaturated ring around the
center of the PSF.

Through these fits, we deduce distances of $6\rlap.{''}75$ and
$12\rlap.{''}92$ from the source to the two knots along the jet.
The same positions are recovered in the two
exposures to within $0\rlap.{''}01$. However, while the first
knot shows a well defined peak, the second knot (further
away from DG Tau) shows a diffuse structure with a width
of $\sim 3''$. Because of this lack of a well defined
peak, we do not use the second knot for calculating
proper motions (see section 3).

\section{Comparing optical, radio, and X-ray knots in DG Tau}

In Table 2, we give the positions of knots along the HH 158
jet compiled by Pyo et al. (2003), together with the positions
of the radio continuum knots seen in Figs. 2 and 3, the position
of the knot closest to the source in our optical images (see \S 2.2)
and the X-ray knot positions of G\"udel et al. (2008, 2011).

We have carried out a least squares fit to the knot positions $x$ as
a function of the time $t$ of the observations of the form:
\begin{equation}
x=x_0+v_T(t-n\,p)\,,
\label{xt}
\end{equation}
where the values of $x_0$, $v_T$ and $p$ are the same for all of the
observed knots, and $n$ is allowed to have values $n=0,$ 1, 2 and 3
(choosing for the individual knots the values of $n$ that minimize
the $\chi^2$). This fit assumes ballistic (i.e. constant velocity)
motions. From the fit, we obtain
\begin{eqnarray}
 p=(4.80\pm 0.32)\,{\rm yr};\,\,
x_0=(550\pm 24)'';\,\, \nonumber 
\end{eqnarray}
\begin{eqnarray}
 v_T=(0.277\pm 0.012)''/{\rm yr}=(198\pm 9)\,{\rm km\,~s^{-1}}
\label{fit}
\end{eqnarray}
where the velocity in km s$^{-1}$ was calculated assuming
a distance of 150~pc to HH 158.
This functional form is appropriate for a system of knots
that travel with a constant velocity $v_T$ (in $''/yr$ for $x$ in
arcseconds and $t$ in years), with an ejection period $p$. The
values of $n$ correspond to the successive knots.

\begin{figure}
   \centering
   \includegraphics[width=8cm]{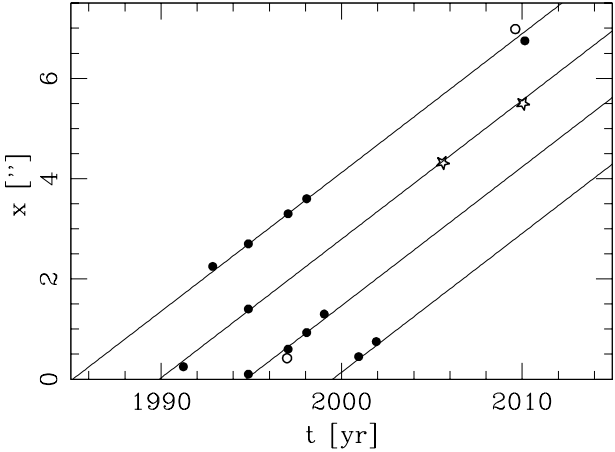}
 \caption{\small Positions of the knots along the HH 158 jet
as a function of time. The filled circles indicate the observed optical
positions, while the open circles and the stars
indicate the observed radio and X-ray knot positions, respectively.
The four straight lines represent
the least squares fit described in section 3 (the four lines
corresponding to the values $n=0$, 1, 2 and 3, from
left to right).}
 \label{knots}
\end{figure}

From the least squares fit to the observed knot positions, we
divide the observed knot positions into four sequences (k0, k1,
k2 and k3, corresponding to $n=0$, 1, 2 and 3, see Table 2).
Fig. \ref{knots} shows the four resulting linear $x$ vs. $t$
dependencies (corresponding to $n=0$, 1, 2 and 3)
together with the observed knot positions.

It is clear that the observed knot positions can be
interpreted as four successive knots, ejected with
a period $p=4.80$~yr and travelling away from the source
at a velocity $v_T=198$~km~s$^{-1}$ (see Eq. \ref{fit}
and Fig. \ref{knots}). This velocity is in good agreement
with previously measured proper motions in HH 158
(see, e. g., Dougados et al. 2000).

\begin{table}[htbp]
\small
      \caption{Optical, radio and X-ray knot positions}
        \begin{center}
            \begin{tabular}{lccccc}\hline\hline
& \multispan4{\phantom{0000000}distance\phantom{0}['']} & \\

Obs. time  & k0 & k1 & k2 & k3 & ref. \\
\hline

1991.24 & \ldots & 0.25 & \ldots & \ldots & 1 \\
1992.86 & 2.25 & \ldots & \ldots & \ldots & 2 \\
1994.84 & 2.7\phantom{0} & 1.4\phantom{0} & 0.1\phantom{0} &
\ldots & 3 \\
1996.98 & \ldots & \ldots & 0.42 & \ldots & ra \\
1997.04 & 3.3\phantom{0} & \ldots & 0.6\phantom{0} & \ldots & 4 \\
1998.07 & 3.6\phantom{0} & \ldots & 0.93 & \ldots & 5 \\
1999.04 & \ldots & \ldots & 1.3\phantom{0} & \ldots & 6 \\
2000.95 & \ldots & \ldots & \ldots & 0.45 & 7 \\
2001.92 & \ldots & \ldots & \ldots & 0.75 & 8 \\
2005.6\phantom{0} & \ldots & 4.32 & \ldots & \ldots & 9 \\
2009.62 & 6.98 & \ldots & \ldots & \ldots & ra \\
2010.05 & \ldots & 5.5 & \ldots & \ldots & 10 \\
2010.15 & 6.75 & \ldots & \ldots & \ldots & op \\

  \hline\hline
\\
\multispan6{ra: radio, this paper\hfil} \\
\multispan6{op: optical, this paper\hfil} \\
\multispan6{1: optical, Kepner et al. (1993)\hfil} \\
\multispan6{2: optical, Solf \& B\"ohm (1993)\hfil} \\
\multispan6{3: optical, Lavalley et al. (1997)\hfil} \\
\multispan6{4: optical, Dougados et al. (2000)\hfil} \\
\multispan6{5: optical, Lavalley et al. (2000)\hfil} \\
\multispan6{6: optical, Bacciotti et al. (2000)\hfil} \\
\multispan6{7: optical, Takami et al. (2002)\hfil} \\
\multispan6{8: optical, Pyo et al. (2003)\hfil} \\
\multispan6{9: X-ray, G\"udel et al. (2008)\hfil} \\
\multispan6{10: X-ray, G\"udel et al. (2011)\hfil} \\
      \label{tab:2}
  \end{tabular}
    \end{center}
\end{table}

From Fig. 5 it is clear that the two radio knots reported by us are
spatially correlated with optical knots that were observed close in time
to the radio observations. On the other hand, the two observations of the X-ray knot
(G\"udel et al. 2008; 2011) show that it is the same knot that
was observed as an optical knot close to the star in the early 1990's
and has displaced several arc seconds since then.
However, in the 2010.05 X-ray observations the knot does not show
optical or radio counterparts that should have beeb seen in the 2010.15 and 2009.62
observations, respectively, presented by us. This lack of counterparts adds to the 
puzzling nature of the X-ray knot in HH~158, whose emission requires models
with shock velocities between 400 and 500 km s$^{-1}$ (G\"unther et al. 2009)
that do not appear to be present from data at other wavelengths.

\section{A model for the radio-continuum emission from
a bipolar outflow}

Raga et al. (1990) developed analytic and numerical models for jets
from sources with intrinsically time-dependent ejection velocities.
These authors showed that supersonic variabilities in the injection
velocity in a supersonic flow result in the formation of two-shock
structures (called working surfaces) that travel down the jet. More
recently, Cant\'o, Raga $\&$ D'Alessio (2000) developed a method
for solving the equations of a supersonic outflow with time-dependent
parameters (ejection velocity and mass loss rate), based on
considerations of momentum conservation for the internal working
surfaces. In particular, these authors obtained solutions for a
sinusoidal velocity variability with constant mass injection rate,
and with constant injection density. In another work, Gonz\'alez $\&$
Cant\'o (2002) developed a model to explain the observed free-free
emission at radio frequencies around low mass stars. In their model,
the ionization is produced by internal shocks in a spherical wind
themselves produced by periodic variations of the injection velocity.
In this section, we present a model for calculating the radio-continuum
(free-free) emission from a stellar flow with conical symmetry. We
assume that the radiation is produced by internal working surfaces
which move inside the bipolar outflow.

\subsection{Dynamics of a working surface}

We consider an outflow which is expelled with constant mass injection
rate $\dot m$, and with an injection velocity $v_e(\tau)$ of the form,
\begin{eqnarray}
v_e(\tau)= v_w - v_c\, \mbox{sin}(\omega \tau),
\label{ve}
\end{eqnarray}
\noindent
where $v_w$ is the mean velocity of the flow, $v_c$ is the amplitude
of the velocity variation and $\omega$ is the angular frequency of
the variation. 

From the formalism developed in Cant\'o et al. (2000), it can be
shown that the first working surface (in each cone) is formed at
a distance $r_c$ from the source given by,
\begin{eqnarray}
r_c= -{{v_w}\over{\omega}}~{{[1-(v_c/v_w)~\mbox{sin}(\omega\tau_c)]^2}
        \over{(v_c/v_w)~\mbox{cos}~\omega\tau_c}},
\label{rc}
\end{eqnarray}
\noindent
at a time,
\begin{eqnarray}
t_c= \tau_c - {{1}\over{w}}~{{[1-(v_c/v_w)~\mbox{sin}(\omega\tau_c)]}
        \over{(v_c/v_w)~\mbox{cos}(\omega\tau_c)}},
\label{tc}
\end{eqnarray}
\noindent
where $\tau_c$ is the ejection time,
\begin{eqnarray}
\tau_c= {{\pi}\over{\omega}}-{{1}\over{\omega}}~
        \mbox{sin}^{-1}~\biggl[ {{-1 + \sqrt{1+ 8(v_c/v_w)^2}}
        \over{2(v_c/v_w)}} \biggr].
\label{tauc}
\end{eqnarray}

Defining the variables
$\overline{\tau}=(\tau_1+\tau_2)/2$ and \boldmath $\Delta \tau=(\tau_2-\tau_1)/2$, 
being $\tau_1$ and $\tau_2$
the ejection time of the flow directly downstream and upstream of
the working surface, respectively, then the equation that describes
the dynamical evolution of the working surface is,
\begin{eqnarray}
a_{\Delta \tau}~\mbox{sin}^2(\omega \bar\tau)+
 b_{\Delta \tau}~\mbox{sin}(\omega \bar\tau) +
 c_{\Delta \tau}= 0,
\label{quad}
\end{eqnarray}
\noindent
where,
\begin{eqnarray}
a_{\Delta \tau}= {{v_c}\over{v_w}}~[(\omega \Delta \tau)^2
 - \mbox{sin}^2(\omega \Delta\tau)],
\label{adel}
\end{eqnarray}
\begin{eqnarray}
b_{\Delta \tau}= (\omega \Delta \tau) \,[\mbox{sin}(\omega
 \Delta\tau) -(\omega \Delta\tau)\,\mbox{cos}(\omega \Delta\tau)],
\label{bdel}
\end{eqnarray}
\begin{eqnarray}
c_{\Delta \tau}= {{v_c}\over{v_w}}~
 \mbox{sin}~(\omega \Delta\tau)\,[\mbox{sin}~(\omega \Delta\tau)
-(\omega \Delta\tau)~\mbox{cos}~(\omega \Delta\tau)
\nonumber
\end{eqnarray}
\begin{eqnarray}
-(\omega \Delta\tau)^2\,\mbox{sin}~(\omega \Delta\tau)].
\label{cdel}
\end{eqnarray}
The $+$ sign in the solution of equation (\ref{quad}) gives
the solution with $|\mbox{sin}(\omega \tau)| \le 1$. Using
$\Delta \tau$ as the free parameter (in the interval $[0,\pi/\omega]$),
we obtain $\bar\tau$ by solving equation (\ref{quad}) and then we
find $\tau_1$ and $\tau_2$. We then use the formalism presented
in Cant\'o et al (2000) to calculate the position and velocity
of the working surface.

\begin{figure}
   \centering
   \includegraphics[width=8cm]{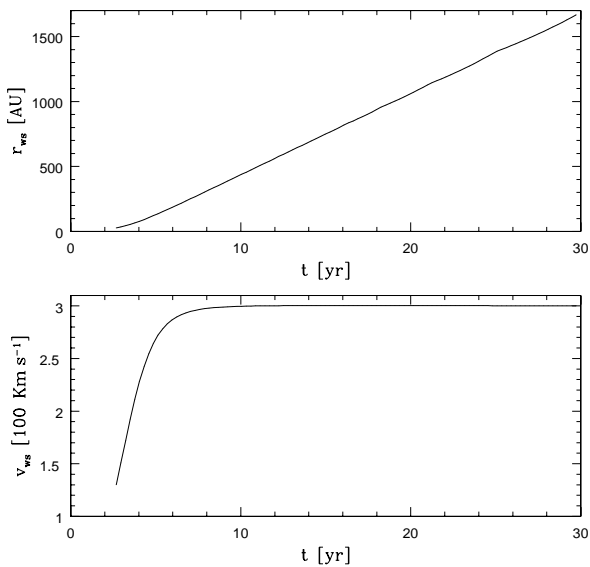}
                 \caption{\small Position $r_{ws}$ and velocity
                  $v_{ws}$ of the working surface formed in a
                  stellar jet with a sinusoidal variation in its
                  velocity at injection. The parameters of the model
                  are $v_{w}= 300$ km s$^{-1}$, $v_{c}= 200$ km s$^{-1}$,
                  and $\omega=$ 1.26 yr. We have assumed a
                  constant mass loss rate of $\dot M= 5 \times 10^{-8}$
                  M$_{\odot}$ yr$^{-1}$. The physical description of the
                  plots is given in the text.}
\label{posvelws}
\end{figure}

In Figure \ref{posvelws}, we show an example
for the position and velocity of the working surface formed in a
stellar jet with a sinusoidal ejection velocity variability. In our model,
we have assumed a constant mass loss rate $\dot M= 5 \times 10^{-8}$
M$_{\odot}$ yr$^{-1}$, mean velocity $v_{w}= 300$ km s$^{-1}$,
amplitude $v_{c}= 200$ km s$^{-1}$, and frequency $w=$ 1.26
yr$^{-1}$ (which corresponds to an oscillation period $P\simeq$
5 yr). It can be observed from the figure that the working
surface is not formed in the flow instantaneously, but at a
time $t_c= 2.67$ yr and at a distance $r_c= 26.81$ AU from the
central star. Later, the working surface begins to be accelerated
asymptotically approaching the velocity $v_w$.

\subsection{The geometric model}

\begin{figure}
   \centering
   \includegraphics[width=8cm]{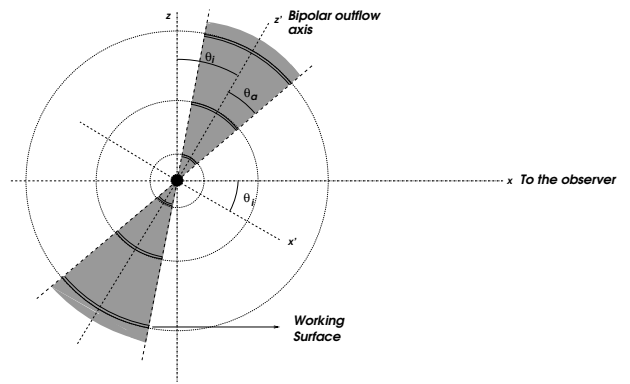}
                 \caption{\small Schematic diagram showing
                 a bipolar ejection with conical symmetry
                 from a central star. The opening angle
                 of the cones is $\theta_a$ and the inclination
                 angle of the outflow axis [$z'$] with
                 the plane of the sky $[y,z]$ is $\theta_i$.
                 The axes $y = y'$ are perpendicular to the
                 figure.}
\label{figjet}
\end{figure}

We consider a stellar outflow expelled from a central star
with a sinusoidal variation in the injection velocity and
constant mass loss rate. We assume conical symmetry for the
jet with an opening angle of the cones $\theta_a$, and an
inclination angle between the outflow axis and the plane of
the sky $\theta_i$. Figure \ref{figjet} shows a schematic diagram
of the bipolar ejection.

A periodical variation in the ejection velocity (eq. [\ref{ve}])
generates a set of outgoing working surfaces. Assuming that the
working surfaces do not lose mass sideways, we can apply the
results presented in section 3.1 to obtain their dynamical
properties (position and velocity as functions of time).
It is possible to construct analytic (or semi-analytic) models
of working surfaces in variable jets under one of the two
following assumptions:
\begin{itemize}
\item the working surfaces eject sideways all of the material
passing through the two shocks (associated with the working surface),
so that the equation of motion is determined by a ram pressure balance
condition,
\item most of the material going through the shocks stays within
the working surface, so that a center of mass formalism can be
constructed.
\end{itemize}
If one compares predictions from (semi)
analytic models (with these two assumptions)
with axisymmetric numerical simulations, it is clear that the
more realistic, axisymmetric solutions to the Euler equation lie
in between the analytic models (see, e.g., Cabrit \& Raga 2000). In
general, the mass conservation assumption is preferred because it
leads to fully analytic solutions for many forms of the ejection
variability (Cant\'o et al. 2000), the ram pressure balance condition
leads to an ordinary differential equation that normally has to be
integrated numerically (see Raga and Cant\'o 1998). Actually, the
models computed under the two approximations described above do
not have stong differences, so that the choice of one over the
other does not introduce important differences with more
realistic jet models (i.e., axisymmetric numerical simulations). 

Under these assumptions,
the first working surface forms at a time $t_c$, the second one
forms after one period $P$, and so on. In this case, the equation
that describes the dynamical evolution of the working surfaces is,
\begin{eqnarray}
\mbox{sin}~[(2k + 1)~\pi-\omega \bar \tau]=
{{-b_{\tiny \Delta \tau} + (b_{\tiny \Delta \tau}
        ^2 -4~ a_{\tiny \Delta \tau} c_{\tiny \Delta \tau})^{1/2}}
        \over{2~ a_{\tiny \Delta \tau}}},
\label{sol2equ2}
\end{eqnarray}
where $k$ is a non-negative integer (0, 1, 2,...). The solution
of equation (\ref{sol2equ2}) with $k= 0$ gives the dynamics of
the first working surface, and, in general, the solution with
$k= j-1$ gives the dynamical evolution of the $j$-th working
surface.

In order to obtain the flux density from the bipolar outflow,
we must find the conditions that indicate us if a working surface
is intersected or not by a given line of sight. In our simplified model,
we assume that every working surface can be described as a portion
of a sphere (a polar cap) whose physical size depends on the opening
angle $\theta_a$ and its position from the central source.
Thus, let $[x,y,z]$ and $[x',y',z']$ be two frames of reference
shown in Figure \ref{figjet}. Every point of the $j$-th
working surface (at a distance $r_{ws[j]}$ from the central star)
satisfies the equation of a sphere in both reference systems, that is,
\begin{eqnarray}
r^2_{ws[j]}= x'^2+y'^2+z'^2= x^2+y^2+z^2,
\label{equrwsjeng}
\end{eqnarray}
and, therefore,

\begin{eqnarray}
x= \pm\, {(r^2_{ws[j]}-y^2-z^2)}^{1/2},
\label{xeng}
\end{eqnarray}
where the symbol $\pm$ indicates the possibility that a line of
sight intersects twice the working surface.

Therefore, the transformation equations between the two frames of
reference are,

\begin{eqnarray}
x'= \pm {(r^2_{ws[j]}-y^2-z^2)}^{1/2}~\mbox{cos}~\theta_i -
 z~\mbox{sin}~\theta_i,
\nonumber
\end{eqnarray}
\begin{eqnarray}
y'= y,
\nonumber
\end{eqnarray}
\begin{eqnarray}
z'= \pm {(r^2_{ws[j]}-y^2-z^2)}^{1/2}~\mbox{sin}~\theta_i +
 z~\mbox{cos}~\theta_i.
\label{tracooreng}
\end{eqnarray}

The intersection conditions of the $j$-th working surface, formed in
the approaching ($z'> 0$) cone or the receding ($z'< 0$) cone,
are obtained by comparison of equation \ref{tracooreng} with the edges
of the caps ($\pm r_{ws[j]}~ \mbox{cos}~ \theta_a$). As a consequence, the $j$-th
working surface is intersected by a given line of sight when,
\begin{eqnarray}
z'\ge r_{ws[j]}~\mbox{cos}~ \theta_a\,,
\label{c1eng}
\end{eqnarray}
or,
\begin{eqnarray}
z'\le - r_{ws[j]}~\mbox{cos}~ \theta_a.
\label{c2ceng}
\end{eqnarray}

Assuming that the observer is located at a distance $D$ from the
source, a given line of sight intersects the plane of the sky
$[y,z]$ at the point ($D\,\mbox{sin}\,\Theta\,\mbox{sin}\,\Phi,
D\,\mbox{sin}\,\Theta\,\mbox{cos}\,\Phi$), where $\Theta$ and
$\Phi$ are the inclination and the azimuthal angles, respectively.
From equations (\ref{tracooreng})-(\ref{c2ceng}), we obtain 
the intersection conditions in terms of these new variables,

\begin{eqnarray}
\pm {(r^2_{ws[j]}-D^2~\mbox{sin}^2 \Theta~ \mbox{sin}^2 \Phi-
                  D^2~ \mbox{sin}^2 \Theta~ \mbox{cos}^2 \Phi)}^{1/2}
~\mbox{sin}~\theta_i
\nonumber
\end{eqnarray}

\begin{eqnarray}
+~ D~ \mbox{sin}\Theta~\mbox{cos}\Phi~\mbox{cos}~\theta_i~ \ge~
 r_{ws[j]}~\mbox{cos}~ \theta_a\,,
\label{cint1}
\end{eqnarray}
or,
\begin{eqnarray}
\pm {(r^2_{ws[j]}-D^2~\mbox{sin}^2 \Theta~ \mbox{sin}^2 \Phi-
                  D^2~ \mbox{sin}^2 \Theta~ \mbox{cos}^2 \Phi)}^{1/2}
 ~\mbox{sin}~\theta_i
\nonumber
\end{eqnarray}

\begin{eqnarray}
+~ D~ \mbox{sin}\Theta~ \mbox{cos}\Phi~\mbox{cos}~\theta_i~ \le~ -
 r_{ws[j]}~\mbox{cos}~ \theta_a\,,
\label{cint2}
\end{eqnarray}
where we have assumed that the observer is far enough that all
points of the polar caps are located at the same distance from the
observer ($D\gg r_{ws[j]}$).

\subsection{Predicted radio-continuum emission}

We consider the model described in section
3.2. First, we add the optical
depths of the working surfaces intersected by each line of sight
to obtain the total optical depth along this line of sight. Then, we
use this optical depth to estimate the intensity emerging from this
direction. Finally, the total flux emitted by the system can be estimated
by integrating this intensity over the solid angle.

Using the numerical models developed by Ghavamian $\&$ Hartigan (1998)
for the free-free emission for a planar interstellar shocks, Gonz\'alez
$\&$ Cant\'o (2002) estimate the average optical depth of a shock wave.
Assuming an average excitation temperature of $10^4$K, these authors
found that their results can be represented by $\tau_{\nu}= \beta\,
n_{0}\,\upsilon_{s}^{\gamma}\,\nu^{-2.1}$, where $n_0$ is the preshock
density, $\upsilon_s$ the shock velocity, and $\nu$ is the frequency.
The constants $\beta$ and $\gamma$ depends on the shock speed.
We note that the optical depth of each working surface has the
the contribution of the internal and external shocks. Using this
representation, the optical depth of the $j$-th working surface is given
by,
\begin{eqnarray}
\tau_{ws[j]}=
        \biggl[ \beta_{e}~ \upsilon^{\gamma_e}_{es}(t_j)~ n_{0,1} (t_j)
                + \beta_{i}~ \upsilon^{\gamma_i}_{is} (t_j)~ n_{0,2} (t_j)
                 \biggr]~ \nu^{-2.1},
\label{tauwsjbipa}
\end{eqnarray}
where $n_{0,1}$ is the preshock density of the external
shock, and $n_{0,2}$ is the preshock density of the internal shock
at its time of dynamical evolution $t_j= t - (j-1)P$.

At a given line of sight (specified by the angles $\theta$ and $\Phi$),
we add the contribution of the $i$-th intersected working surfaces to obtain
the total optical depth $\tau_{\nu}(\Theta, \Phi)$ along this line of sight,
that is,
\begin{eqnarray}
\tau_{\nu}(\Theta, \Phi)= \sum_{j=1}^i~
        {{\tau_{ws[j]}~(\Theta, \Phi)} \over
                {\mu_j}},
\label{sinsumaibip}
\end{eqnarray}
where $\mu_j= \mbox{cos}\,\theta_j$, being $\theta_j$ the angle
between the line of sight and the normal vector to the $j$-th working
surface at the intersection point. It is easy to show that $\mu_j$
can be written as,

\begin{eqnarray}
\mu_j= \biggl [ 1 + {{\tilde r^2_{ws[1]}}
                                \over{\tilde r^2_{ws[j]}}}
                ~(\mu_1-1) \biggr]^{1/2}\,.
\nonumber
\end{eqnarray}

Finally, the flux density at radio frequencies from the bipolar
outflow can be calculated by,

\begin{eqnarray}
S_{\nu}= B_{\nu}~ \int_0^{2\pi} \int_0^{\Theta_c}
                 \biggl(1 - e^{-\tau_{\nu}(\Theta, \Phi)}
                        \biggr)~\mbox{sin}~\Theta~d\Theta~d\Phi,
\label{sinfluxbip}
\end{eqnarray}
where $B_{\nu}$ is the Planck function in the Rayleigh-Jeans
approximation ($B_{\nu}= 2kT_e\,\nu^2/c^2$ being $k$ the Boltzmann
constant, $T_e$ the electron temperature and $c$ the speed of light),
and $\Theta_c= \mbox{atan}\, (r_{ws[1]}/D)$.

\begin{figure}
   \centering
   \includegraphics[width=8cm]{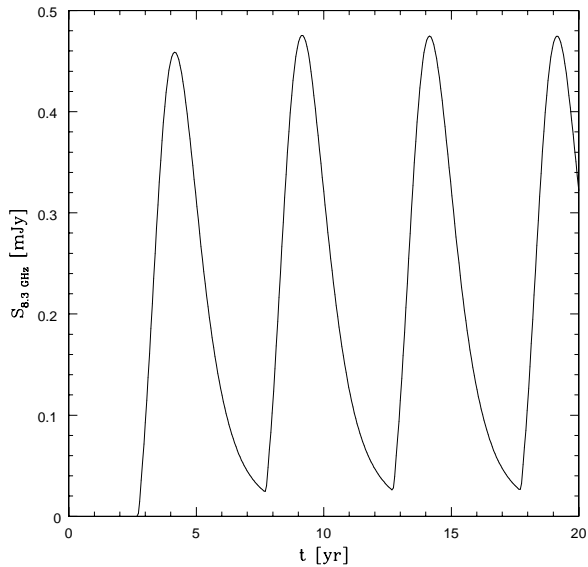}
                 \caption{\small Predicted radio-continuum
                 fluxes at $\lambda=$ 3.6 cm for a
                 bipolar outflow with a sinusoidal ejection
                 velocity. The assumed parameters are
                 $v_w$= 300 km s$^{-1}$, $v_{c}= 200$ km s$^{-1}$,
                 and $\omega= 1.26$ yr$^{-1}$.
                 We have adopted a constant mass loss rate
                 $\dot m= 5\times10^{-8}$ M$_{\odot}$ yr$^{-1}$,
                 and a distance $D=$ 150 pc from the observer.
                 The opening angle of the cones are $\theta_a=
                 30^{\sf o}$ with an inclination angle $\theta_i=
                 42^{\sf o}$. The physical description of the plot
                 is given in the text.}
\label{fluxjet}
\end{figure}

\subsection{Numerical example}

In this section, we present a numerical example for the
predicted radio-continuum flux at $\lambda=$ 3.6\,cm from
a bipolar outflow with a sinusoidal ejection velocity. The
opening angle of the cones is $\theta_a= 30^{\sf o}$ and the
inclination angle between the outflow axis and the sky plane is
$\theta_i= 42^{\sf o}$. The outflow is ejected from the central
star with a mean velocity $v_w$= 300 km s$^{-1}$, an amplitude
$v_{c}= 200$ km s$^{-1}$ and with an oscillation period
$P= 5$ yr ($\omega=$ 1.26$\,yr^{-1}$). We have assumed a constant
mass loss rate $\dot m= 5\times 10^{-8}$ M$_{\odot}$ yr$^{-1}$
and a distance $D=$ 150 pc from the observer. These values
are consistent with the parameters estimated from HH 158 and other
jets from low-mass young stars (Eisloffel et al. 2000; Pyo et al. 2003; Agra-Amboage et 
al. 2011).

In Figure \ref{fluxjet}, we present our results. Once the
first working surfaces (in both cones) are formed, at a time
$t= 2.67$\,yr (see, also, Fig. \ref{posvelws}), the flux increases
reaching a maximum value of $S_{8.46~GHz}\sim$\, 0.5\,mJy at
$t\simeq$\,4 yr. Afterwards, the emission decreases until
new working surfaces are formed in the outflow, and the flux
increases again. We note that the flux density shows a periodical
behaviour with the same period as that of the injection velocity variability.
Finally, we note from the figure that the predicted values of
the flux density are in agreement with the radio observations
of HH 158 reported in section 2.

\section{Summary and Conclusions}

We presented an analysis of archive as well as new Very Large Array observations
of DG Tau that detect emission from knots in the jet associated with this star.
Radio knots are detected in the 1996.98 and 2009.62 data and they are found to correlate
with optical knots in observations made close in time to the radio observations.
One of these optical observations is provided by us in this paper. In contrast, 
the X-ray knot that is observed in the G\"udel et al. (2011) data does not
coincide with the radio/optical knot and appears to be part of a different, later ejection that
was first detected as a moving optical knot in the early 1990's.

All the observed knot positions (optical, radio, and X-ray) can be
interpreted as four successive knots, ejected with
a period $p=4.80$~yr and travelling away from the source
at a velocity $v_T=198$~km~s$^{-1}$. The next knot ejection is expected to
take place around epoch 2014.0.

We have modeled successfully the observed radio continuum emission in terms of
working surfaces produced in a jet with a velocity at injection that varies
sinusoidally with time.  

\begin{acknowledgements}
The National Radio
Astronomy Observatory is a facility of the National Science Foundation
operated under cooperative agreement by Associated Universities, Inc.
The paper is partially based on observations made with the Nordic
Optical Telescope, operated on the island of La Palma jointly by
Denmark, Finland, Iceland, Norway, and Sweden, in the Spanish
Observatorio del Roque de los Muchachos of the Instituto de
Astrof\'\i sica de Canarias. LFR, AGR, JC, LL and LZ
acknowledge the financial support of DGAPA, UNAM and CONACyT, 
M\'exico.
RFG ackowledges support from grant PAPIIT IN100511-2, UNAM, M\'exico.
LL is indebted to the Alexander von Humboldt Stiftung and the Guggenheim Memorial 
Foundation  for financial support.

\end{acknowledgements}

\end{document}